\documentclass[runningheads]{llncs}
\usepackage{graphicx}
\usepackage{amsfonts}
\usepackage{amssymb}
\usepackage{hyperref}

\usepackage[utf8]{inputenc} 

\usepackage{soul}

\usepackage{lipsum}
\usepackage{blindtext}

\usepackage{makecell}
\usepackage{booktabs,tabularx}
\usepackage{graphicx}
\usepackage{pdflscape}

\usepackage{multirow}

\newcommand{\et}{et al.\xspace}

\begin{document}
%
\title{Network-Based Delineation of Health Service Areas: A Comparative Analysis of Community Detection Algorithms}
%
\titlerunning{Network-Based HSA Delineation: A Comparative Analysis}
%
\author{Diego Pinheiro\inst{1}\orcidID{0000-0001-9300-7196} \and
Ryan Hartman\inst{3}\orcidID{0000-0001-8044-5461} 
\and
Erick Romero\inst{1}\orcidID{0000-0001-5456-3093} 
\and
Ronaldo Menezes\inst{2}\orcidID{0000-0002-6479-6429}
\and
Martin Cadeiras\inst{1}\orcidID{0000-0003-4545-2871}
}
\authorrunning{Pinheiro et al.}
%
\institute{Department of Internal Medicine, University of California, Davis, USA \and
Department of Computer Science, University of Exeter, UK
\and
Independent Researcher\\
}
\maketitle              
\begin{abstract}
A Health Service Area (HSA) is a group of geographic regions served by similar health care facilities. 
The delineation of HSAs plays a pivotal role in the characterization of health care services available in an area, enabling a better planning and regulation of health care services.
Though Dartmouth HSAs have been the standard delineation for decades, previous work has recently shown an improved HSA delineation using a network-based approach, in which HSAs are the communities extracted by the Louvain algorithm in hospital-patient discharge networks.
Given the existent heterogeneity of communities extracted by different community detection algorithms, a comparative analysis of community detection algorithms for optimal HSA delineation is lacking.
In this work, we compared HSA delineations produced by community detection algorithms using a large-scale dataset containing different types of hospital-patient discharges spanning a 7-year period in US.
Our results replicated the heterogeneity among community detection algorithms found in previous works, the improved HSA delineation obtained by a network-based, and suggested that Infomap may be a more suitable community detection for HSA delineation since it finds a high number of HSAs with high localization index and a low network conductance. 

\keywords{Hospital-Patient Discharge Networks \and Community Detection Algorithms \and Health Service Area  \and Comparative Analysis}
\end{abstract}

\section{Introduction}

A Health Service Area (HSA) is as a group of geographic regions in which residing patients most often receive healthcare services from similar health care facilities. It was first introduced in 1973 by Wennberg and Gittelsohn as a more meaningful unit of analysis for healthcare data than administrative geographic divisions such as counties~\cite{Wennberg:1973fl}. By examining variations in expenditures among HSAs delineated in Vermont, the authors showed, for instance, that the expenditure per capita among HSAs varied from \$54 to \$162 and such variation, however, had no correlation with age-adjusted mortality. 

In 1996, Wenneberg then proposed the {\it Dartmouth Atlas of Health Care in the United States}~\cite{Wennberg:1996th} which is the current standard HSA delineation in the US. Effectively, Each Dartmouth-HSAs are delineated in three steps. First, each health care facility is assigned to its respective city/town. Then, each ZIP Code is assigned to the city/town of the health care facility from which residing patients receive most of their healthcare services. As a result, each Dartmouth-HSA is the group of ZIP Codes associated to the same city/town. Finally, enclave ZIP Codes, if any, are assigned the city/town of its adjacent ZIP Codes to unsure geographic contiguity.  

Recently, Hu \et~\cite{Hu:2018jl} has proposed a network-based approach to HSA delineation in which a Hospital-Patient Discharge Network (HPDN) was built and the community detection algorithm Louvain was subsequently applied to find communities (i.e., HSAs) with the highest network modularity. In their HPDN, nodes represent distinct ZIP Codes and links represent the total number of discharges between the ZIP Codes of health care facilities and patient residencies. Using claims-based hospital discharges in Florida, the authors demonstrated that Louvain-HSAs presented, for instance, a higher localization index than Dartmouth-HSAs, which is measure of internal validity that quantifies the proportion of patients receiving services from health care facilities located at the same HSA in which they reside.

Yet, a comparative analysis of community detection algorithms, is still lacking for optimal network-based HSA delineation. Such comparative analysis is needed given the heterogeneity of communities extracted by different community detection algorithms~\cite{Hartman:2017go}. Though such comparative analysis was previously provided for grouping hospitals~\cite{Everson:2019kh}, the underlying networks differ from HPDNs in two fundamental aspects: nodes were hospitals instead of geographical regions, and links were patients sharing between hospitals instead of the total number of hospital discharges. 

In this paper, a comparative analysis of was conducted for HSAs delineated by four commonly used community detection algorithms, namely, Block Model~\cite{Peixoto2014}, Infomap~\cite{Rosvall2008} , Louvain~\cite{Blondel:2008do}, and Speaker-Listener Label Propagation Algorithm (SLPA)~\cite{Xie2011}, were compared. A claims-based patient-hospital discharge data was used; it containing a total of $124,970,471$ discharges over a 7-year period in California, US. Our results replicated the existent heterogeneity of communities extracted by different algorithms of community detection and reinforced the use of a network-based approach to HSA delineation. The results demonstrated, for instance, that Infomap was the most suitable algorithm because it was capable of delineating a high number of HSAs, and Infomap-HSAs still presented a high localization index and a low network conductance. To shift the standard HSA delineation methodology from Dartmouth to a network-based approach, further studies on the reliability, validity, and generality of a network-based approach are required.

\section{Methodology}

\begin{figure}[ht!]
	\centering
	\includegraphics[width=\textwidth]{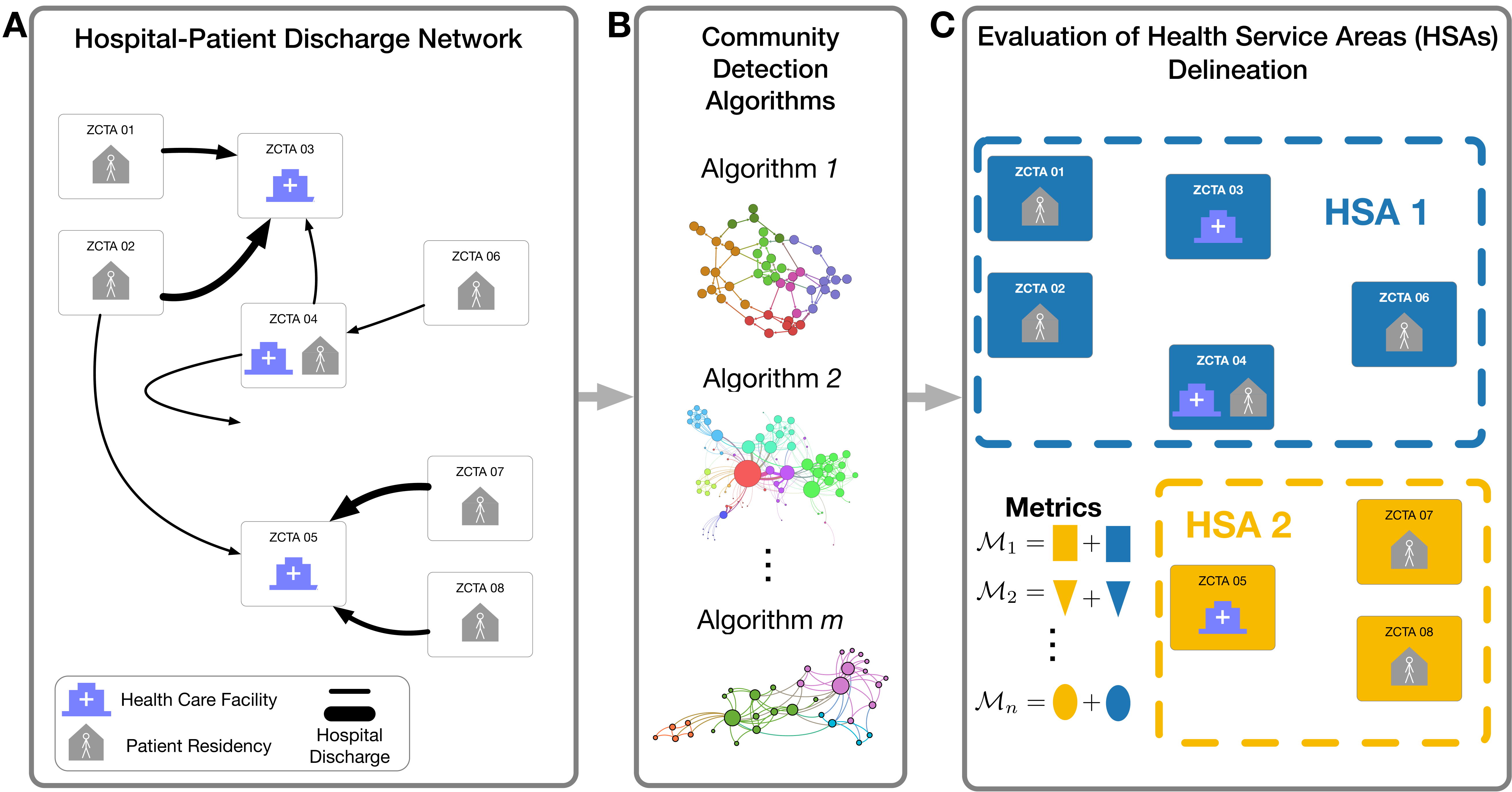}
	\caption{Network-based delineation of Health Service Areas (HSAs). (\textbf{A}) Hospital discharge data is used to build Hospital-Patient Discharge Networks (HPDN) in which nodes represent ZIP Code Tabulation Areas (ZCTAs), and links represent the total number of discharges between the ZCTA locations of health care facilities and patient residencies. (\textbf{B}) Community detection algorithms are applied over HDPNs to delineate HSAs. (\textbf{C}) Delineated HSAs are evaluated according to multiple quality metrics of HSA delineation.}
	\label{fig:diagram}
\end{figure}

A network-based delineation of Health Service Areas (\figurename~\ref{fig:diagram}) consists of (A) modeling claims-based patient-hospital discharge data as networks of ZIP Code Tabulation Areas (ZCTAs), (B) applying a diverse set of community detection algorithms, and (C) performing a comparative analysis of the extracted communities (i.e., HSAs) according to multiple quality metrics of HSA delineation. Overall, $28$ Hospital-Patient Discharge Network (HPDN) were built, one for each combination of $4$ discharge types and  $7$ years. For each HPDN, each of the $4$ community detection algorithms was applied, and a total of $112$ HSA delineations were obtained. Though only a subset of HPDNs and HSA delineations are presented, all of the code, datasets, networks, and analysis are available on the Open Science Framework (OSF) repository of this project at \url{https://doi.org/10.17605/OSF.IO/GW73Y}. 

\subsection{Hospital-Patient Discharge Networks}

Hospital-Patient Discharge Networks (HPDNs) were built using claims-based hospital discharge data obtained from the California Health and Human Services Agency (CHHS)~\cite{chhs}. This dataset is publicly available and contains a total of $124,970,471$ hospital-patient discharges of different types spanning a $7$-year period from 2012 through 2018. The four discharge types are {\it Inpatient from Emergency Department (ED)}, {\it Inpatient}, {\it Ambulatory Surgery}, and {\it ED Only}.

Each data point contains the type of discharge, the year, the name of the facility (e.g., Alameda Hospital, University of California Davis Medical Center). Also, it contains the 5-digit ZIP Codes (e.g., 94501, 95831) of both facility location and patient residency as well as the respective number of discharges between them. Hospital discharges to patient residency ZIP Codes other than those with 5-digits were excluded (2.6\%); these were mainly discharges of patients from states other than California (e.g., ARIZONA, NEVADA (state), Other U.S.), from locations outside the US (e.g., OUTSIDE U.S.), from unknown locations (e.g., UNKNOWN), and from homeless population (e.g., HOMELESS). 

ZIP Codes are a collection of delivery routes maintained U.S. Postal Service and ZIP Code Tabulation Area (ZCTAs) are actual generalized areal representations maintained by the U.S. Census Bureau. Therefore, for each ZIP Code of both health care facility and patient residency, the corresponding ZCTA was obtained using the ZIP Code to ZCTA Crosswalk provided from the Uniform Data System Mapper~\cite{udsmapper}. 

A separate weighted and undirected HPDN network was built for each type of hospital discharge and for each year according to the methodology proposed by Hu \et~\cite{Hu:2018jl}. In each network, nodes are ZCTAs, links are the total number of discharges between the ZCTAs of health care facilities and patient residencies. The HPDN is an undirected network because a link $w_{ij}$ encodes the total number of discharges {\it between} ZCTAs $i$ and $j$ without arbitrarily distinguishing whether the direction is due to a health care facility at $i$ discharging patients residing at $j$ or patients residing at $i$ going to a facility at $j$ for health care services.  

\subsection{Community Detection}

In a network-based HSA delineation, each HSA correspond to a distinct community extracted by a community detection algorithm from a hospital-patient discharge network (HPDN)~\cite{Hu:2018jl}. While the existence of communities in real-world networks is agreed upon, there is no generally accepted definition of what a community is, or what the most appropriate way to find them is~\cite{Fortunato:2016if}.  Some algorithms take a stronger approach to community detection by looking for cliques, which are a group of nodes for which there is a link between every pair of nodes~\cite{Fortunato:2010iw}; other approaches just look for more densely connected subgraphs within the network such that a community is a subset of nodes within a network that are densely connected to each other when compared to the rest of the network.

The lack of a general definition of what a community should be is also mirrored in the existent heterogeneity of communities extracted by different community detection algorithms~\cite{Hartman:2017go} and as such requires a comparison among community detection algorithms. Four commonly used algorithms were selected: Louvain Modularity, Infomap, Stochastic Block Model, and Speaker-listener Label Propagation. These algorithms were selected as they provide four very distinct approaches to community detection and have their implementations easily provided by their authors.

The {\it Louvain} algorithm~\cite{Blondel:2008do} finds communities that maximizes the network modularity. This quantifies the extent to which the density of links within the found communities excessively surpasses that of what would be expected if links were placed at random. Trying all possible partitions is not computationally feasible, and Louvain modularity takes a heuristic approach by maximizing the local modularity of smaller communities that are only subsequently joined if such aggregation leads to an increased modularity. These smaller communities start as individual nodes and are iteratively joined together into greater communities until the a single community containing the whole network is reached. 

The {\it Stochastic Block Model}~\cite{Peixoto2014} uses a maximum likelihood estimator to infer the block structure of the network. This algorithm attempts to recover the hierarchical block structure of the network where each block represents a community. The block model used in this study is the degree-corrected variant given the weighted aspect of HPDNs and that previous work has shown that the such variant tends to perform better on empirical networks~\cite{Peixoto2015}. 

The {\it Infomap} algorithm~\cite{Rosvall2008} finds communities that maximizes the map equation instead of modularity. The map equation quantifies the length of the coding scheme necessary to communicate the sequence of movements of random walkers within the network. In essence, if a community structure exists, random walkers will become trapped within these communities because movements within-communities are more likely than between-communities. Therefore, the coding scheme necessary to communicate the sequence of movements can be reduced by taking into account the community structure as every time a walker enters into a community, the community is identified, and a smaller community-specific coding scheme is used to quantify within-community movements.

The {\it Speaker-Listener Label Propagation Algorithm} (SLPA)~\cite{Xie2011} is a localized community detection algorithm based on the concept of label propagation. SLPA finds communities by initially assigning each node to a unique label. Nodes then iteratively changes their label to the label most often used by its neighbors. Initially, this label exchange rule promotes the formation of smaller consensus groups which will subsequently compete with one another for node members depending on  the balance between their within-group and between-group interactions. In contrast to the other algorithm's, SLPA does not require prior information about the network, nor does it attempt to maximize any metric as a proxy for well-defined communities. Instead, it only relies on the network structure to identify the communities. While SLPA is able to find overlapping communities, the post processing threshold was set to $r=0.5$ to ensure the extraction of non-overlapping communities and thus provide a better comparison to the other non overlapping algorithms used.

\subsection{Evaluation of Health Service Areas (HSA) Delineation}

The quality of a HSA delineation can be evaluated according to multiple and often conflicting metrics ~\cite{Hu:2018jl,Everson:2019kh}. The following four delineation metrics of a HSA $c$ were used: the number of communities ($N_C$), the localization index ($LI_c$),network conductance ($C_c$), and the total number of discharges ($D_c$). 

In a network-based HSA delineation, the {\it total number of delineated HSAs} $N_C$ is determined by the specific community detection algorithm used as it is the number of communities extracted. The $N_C$ is an important metric since it determines the number of distinct meaningful units of analysis ultimately uncovered and, as in any community detection problem, trivial solutions such as $1$ (one) and $n$, the total number of nodes,  are generally undesired~\cite{Hartman:2017go}.

The {\it localization index} $LI_c$ of a community $c$ quantifies the proportion of patients seeking and receiving health care services from hospitals within the HSA in which they reside. A high value of $LI_c$ is desired since it describes HSA $c$ uncovered a geographic region in which residing patients receive comparable health care services. Formally, the localization index $LI_c$ of community $c$ can be defined as
\begin{equation}
 LI_c = \frac{N_{D(c,c)}}{N_{D(c)}} \enspace ,
\end{equation}
in which $N_{D(i,j)}$ is the total number of discharges from patients residing at ZCTAs within community $i$ that are discharged from hospitals at ZCTAs within community $j$, and $N_{D(c)} = \sum_j^{N_C} N_{D(c,j)}$ is the total number of discharges from patients living within community $c$.

The network conductance $C_{c}$ of a community $c$ is a network-based measure which quantifies the extent to $c$ is well-formed. It is calculated using the total links running within its community $i$ relative to links running from $i$ to other communities. Conductance is based on the degree-based definition of a community~\cite{Fortunato:2016if}. Formally, the conductance $C_(c)$ of a community $c$ can be calculated as 
\begin{equation}
C_{c} = \frac{w_C^{ext}}{w_C} \enspace ,
\end{equation}
in which, for a weighted network, $W_{ij}$ is the strength of the link connecting nodes $i$ and $j$, $w_C = \sum_{i \in C, j} W_{ij}$ is the total strength of links originating within community $c$, and $w_C^{ext} = \sum_{i \in C, j \notin C} W_{ij}$ is the external strength.

Aside from the aforementioned metrics, the total number of discharges $N_D(c)$ from patients living in one of the ZCTAs found within community $c$ is also calculated. Ideally, a community detection algorithm would maximize the number of HSAs where each has a high localization index and a low conductance. Yet, as the total number of communities increases from one to $n$, the typical value of localization index decreases from $1$ to $0$ and the typical value of network conductance increases from $0$ to $1$. 

To provide a reliable estimate for each metric, $B = 1,000$ bootstrap samples with replacement draw from the distribution of each metric were used to calculate a mean value for the localization index $\langle li \rangle$, network conductance $\langle c \rangle$, and total number of discharges $\langle d \rangle$. The standard deviation was also provided for each of the aforementioned estimators. 

\section{Results and Discussion}

The network statistics of each individual HPDN varied among discharge types and over the years (\tablename~\ref{tab:net_summary}). Considering the year of 2012, for instance, {\it Inpatient from ED} and {\it ED Only} HPDNs had comparable total number of nodes $n$. Their number of links $m$ were $49,000$ and $127,000$, respectively, suggesting that a {\it ED Only} HPDN has a higher number of distinct ZCTA pairs for which hospital discharges occurred between hospital locations and patient residencies. Also, their the total network link strength $w$ were $1.6$ million and $9.2$ million, respectively, and their network density $\rho$ were $0.0345$ and $0.0883$, respectively,suggesting that the healthcare service underlying a {\it ED Only} HPDN has a higher demand and is more dense. Over the years, the average shortest path length $l$ and clustering coefficient $c$ of HPDNs slightly decreased. 

\begin{table}[!ht]
	\caption{Statistics of Hospital-Patient Discharge Networks (HPDNs). For each type of discharge and for each year, the following network measures were quantified: the total number of nodes ($n$), the total number of links ($m$), the total weight ($w$), the network density ($\rho$), the average shortest path length ($l$), as well as the clustering coefficient ($c$). Other types of discharges and network metrics are available on the Open Science Framework (OSF) repository of this project at \url{https://doi.org/10.17605/OSF.IO/GW73Y}.}
		\label{tab:net_summary}
	\begin{center}
		\tabcolsep=0.15cm
		\begin{tabular}{p{1.4cm}p{.5cm}cccccc}
			\toprule
			&  &      $n$ &      $m$ &      $w$ &   $\rho$ &    $l$ &      $c$ \\
			Type of Discharge & Year &          &          &          &          &        &          \\
			\midrule
			\multirow{7}{*}{\shortstack[l]{Inpatient \\ from ED}} & 2012 & 1.69E+03 & 4.92E+04 & 1.63E+06 & 3.45E-02 & $2.81$ & 1.19E-03 \\
			& 2013 & 1.69E+03 & 4.97E+04 & 1.63E+06 & 3.49E-02 & $2.72$ & 1.12E-03 \\
			& 2014 & 1.69E+03 & 5.05E+04 & 1.64E+06 & 3.56E-02 & $2.71$ & 1.02E-03 \\
			& 2015 & 1.68E+03 & 5.25E+04 & 1.72E+06 & 3.70E-02 & $2.89$ & 9.74E-04 \\
			& 2016 & 1.75E+03 & 5.37E+04 & 1.74E+06 & 3.50E-02 & $2.70$ & 9.53E-04 \\
			& 2017 & 1.75E+03 & 5.38E+04 & 1.75E+06 & 3.52E-02 & $2.67$ & 9.84E-04 \\
			& 2018 & 1.75E+03 & 5.41E+04 & 1.75E+06 & 3.54E-02 & $2.62$ & 9.31E-04 \\
			\cline{1-8}
			\multirow{7}{*}{ED Only} & 2012 & 1.69E+03 & 1.27E+05 & 9.25E+06 & 8.83E-02 & $2.15$ & 2.14E-04 \\
			& 2013 & 1.69E+03 & 1.28E+05 & 9.65E+06 & 8.95E-02 & $2.18$ & 2.07E-04 \\
			& 2014 & 1.69E+03 & 1.33E+05 & 1.03E+07 & 9.25E-02 & $2.20$ & 1.90E-04 \\
			& 2015 & 1.69E+03 & 1.39E+05 & 1.12E+07 & 9.66E-02 & $2.16$ & 1.78E-04 \\
			& 2016 & 1.76E+03 & 1.42E+05 & 1.15E+07 & 9.15E-02 & $2.15$ & 1.81E-04 \\
			& 2017 & 1.76E+03 & 1.43E+05 & 1.17E+07 & 9.25E-02 & $2.15$ & 1.88E-04 \\
			& 2018 & 1.76E+03 & 1.42E+05 & 1.14E+07 & 9.17E-02 & $2.16$ & 1.96E-04 \\
			\bottomrule
		\end{tabular}
	\end{center}
\end{table}

\begin{landscape}
\begin{figure}[t]
    \centering
    \includegraphics[width=.92\columnwidth]{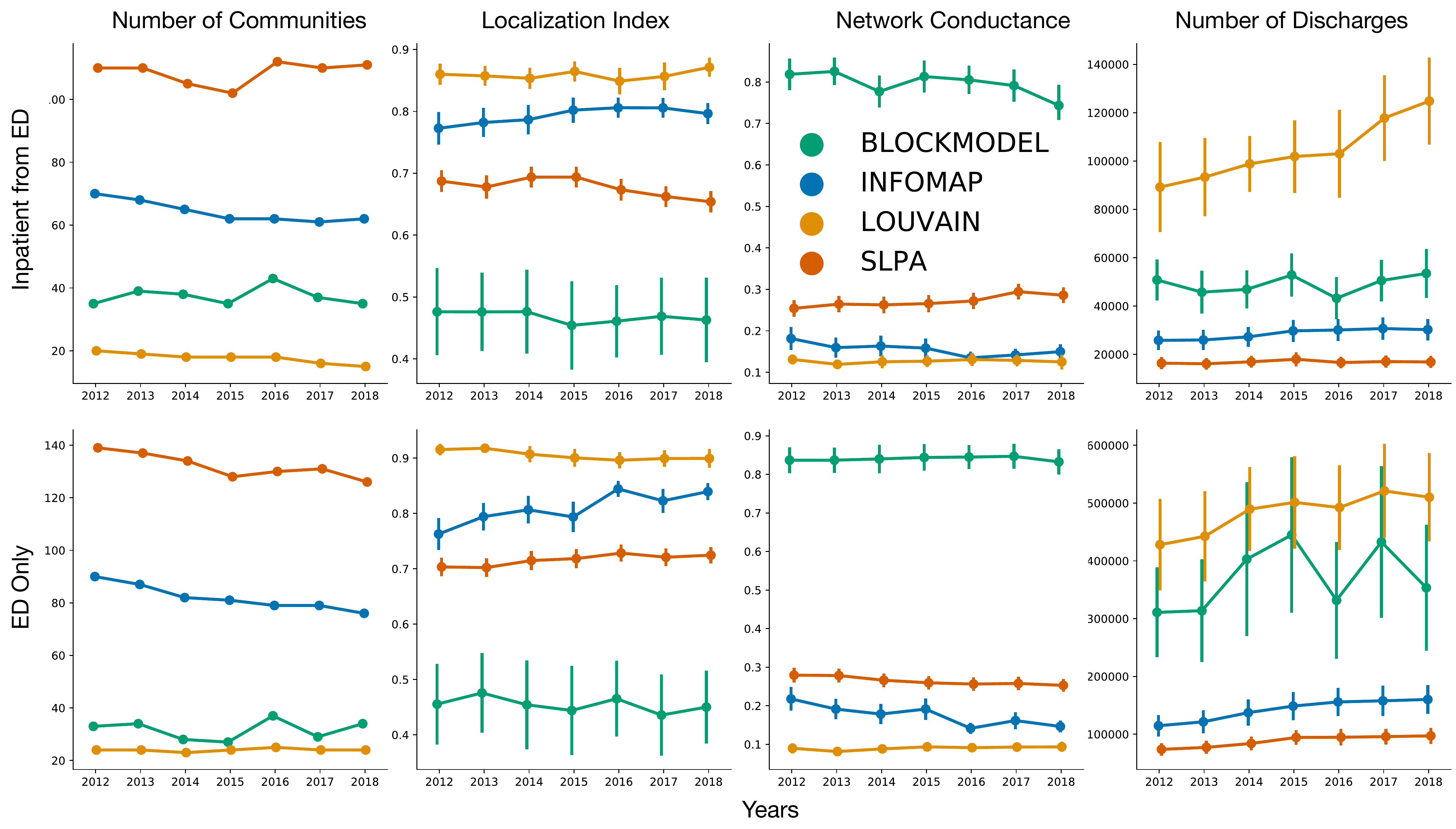}
    \caption{Comparison of HSAs delineated using Block Model, Infomap, Louvain, and SLPA community detection algorithms. The HSA delineations were compared in terms of the number of communities ($N_C$), localization index ($LI(c)$), network conductance ($C_(c)$), and total number of discharges ($N_D(i,j))$. The discharge types presented are {\it Inpatient from ED} (top) and {\it ED Only} (bottom). All results are available on the Open Science Framework (OSF) repository of this project at \url{https://doi.org/10.17605/OSF.IO/GW73Y}.}
    \label{fig:community_size}
\end{figure}
\end{landscape}

Overall, HSA delineation results (\figurename~\ref{fig:community_size} and \tablename~\ref{tab:stats_community_detection}) have reinforced the superiority of network-based HSA delineation as well as confirmed the heterogeneity among communities extracted by different community detection algorithms even in the same dataset. Such results are not only consistent with community detection comparisons from previous works~\cite{Hartman:2017go}, but also advocate for a further comparison among community detection algorithms.

\begin{table}[!ht]
\caption{Comparison of HSAs delineated using the community detection algorithms Block Model, Infomap, Louvain, and SLPA in terms of in terms of the number of communities ($n_c$) as well as the typical values of the localization index $\langle li \rangle$, network conductance $\langle c \rangle$, and total number of discharges $\langle d \rangle$. The discharge types presented are {\it Inpatient from ED} and {\it ED Only} for the years of 2012 and 2018. The other discharge types and years are available on the Open Science Framework (OSF) repository of this project at \url{https://doi.org/10.17605/OSF.IO/GW73Y}.}
	\label{tab:stats_community_detection}
	\tabcolsep=0.17cm
	\begin{tabular}{lllrrrr}
		\toprule
		&      &      & $n_c$ & $\langle li \rangle$ & $\langle c \rangle$ &  $\langle d \rangle$ \\
		Type of Discharge & Year & Community Detection &       &                      &                     &                      \\
		\midrule
		\multirow{8}{*}{Inpatient from ED} & \multirow{4}{*}{2012} & BLOCK MODEL &  $35$ &               $0.47$ &              $0.82$ &             $51,450$ \\
		&      & INFOMAP &  $70$ &               $0.77$ &              $0.18$ &             $25,539$ \\
		&      & LOUVAIN &  $20$ &               $0.86$ &              $0.13$ &             $89,235$ \\
		&      & SLPA & $110$ &               $0.69$ &              $0.25$ &             $16,392$ \\
		\cline{2-7}
		& \multirow{4}{*}{2018} & BLOCK MODEL &  $35$ &               $0.47$ &              $0.74$ &             $54,090$ \\
		&      & INFOMAP &  $62$ &               $0.80$ &              $0.15$ &             $30,397$ \\
		&      & LOUVAIN &  $15$ &               $0.87$ &              $0.13$ &            $125,833$ \\
		&      & SLPA & $111$ &               $0.65$ &              $0.28$ &             $16,830$ \\
		\cline{1-7}
		\cline{2-7}
		\multirow{8}{*}{ED Only} & \multirow{4}{*}{2012} & BLOCK MODEL &  $33$ &               $0.45$ &              $0.84$ &            $311,009$ \\
		&      & INFOMAP &  $90$ &               $0.77$ &              $0.22$ &            $114,117$ \\
		&      & LOUVAIN &  $24$ &               $0.92$ &              $0.09$ &            $430,396$ \\
		&      & SLPA & $139$ &               $0.70$ &              $0.28$ &             $74,049$ \\
		\cline{2-7}
		& \multirow{4}{*}{2018} & BLOCK MODEL &  $34$ &               $0.45$ &              $0.83$ &            $359,908$ \\
		&      & INFOMAP &  $76$ &               $0.84$ &              $0.15$ &            $160,869$ \\
		&      & LOUVAIN &  $24$ &               $0.90$ &              $0.09$ &            $512,044$ \\
		&      & SLPA & $126$ &               $0.73$ &              $0.25$ &             $97,810$ \\
		\bottomrule
	\end{tabular}
\end{table}

The comparison of HSA delineations (\figurename~\ref{fig:community_size}) involves the evaluation of conflicting metrics such as higher number of communities and localization index. Considering the discharge type {\it ED Only} and year 2018 (\tablename~\ref{tab:stats_community_detection}), for instance, the difference between the number Louvain-HSAs and SLPA-HSAs was $4$-fold, with $24$ Louvain-HSAs and $126$ SLPA-HSAs. This fewer number of Louvain-HSAs using hospitals discharge in California is also consistent with the fewer number of Louvain-HSAs in the previous work from Hu~\et~\cite{Hu:2018jl} using hospital discharges in Florida. Though the localization index of Louvain-HSAs ($.90$) was 7\% higher than that of Infomap-HSAs ($.84$), $52$ more Infomap-HSAs were delineated, representing a $2$-fold increase. 

\begin{landscape}
\begin{figure}[t]
    \centering
    \includegraphics[width=.90\columnwidth]{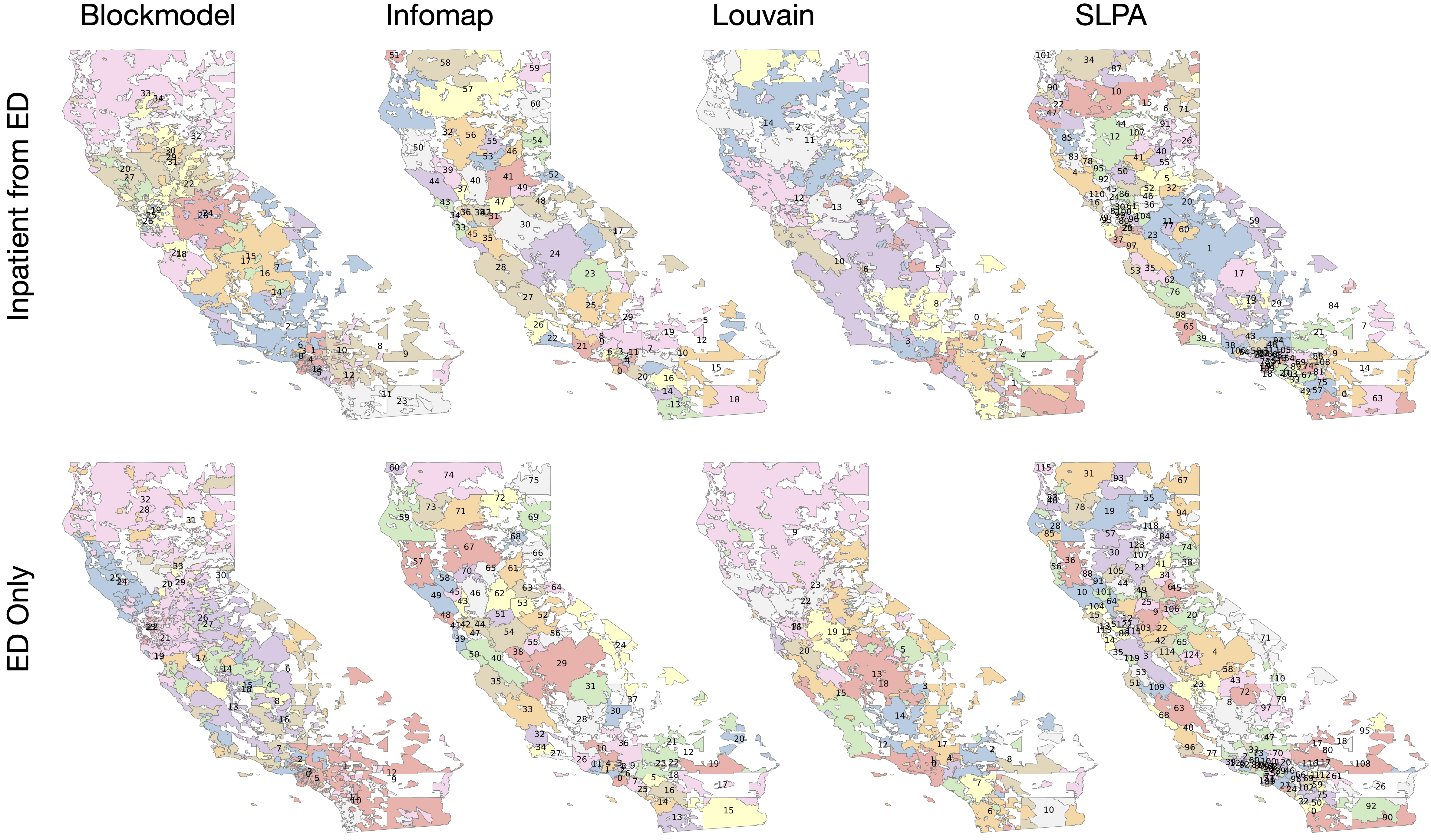}
    \caption{Maps of HSA delineated using the community detection algorithms Block Model, Infomap, Louvain, and SLPA for discharge types {\it Inpatient from ED} (top) {\it ED Only} (bottom) in 2018. Each HSA is displayed as a geographical boundary aggregating one or more ZCTAs along with its respective Id. HSAs were then colored using a color blind palette with 9 distinct colors. All maps are available on the Open Science Framework (OSF) repository of this project at \url{https://doi.org/10.17605/OSF.IO/GW73Y} }
    \label{fig:maps_discharge}
\end{figure}
\end{landscape}

By examining their geographical patterns (\figurename~\ref{fig:maps_discharge}), the respective geographical areas of Louvain-HSAs correspond to a wider and more discontinuous geographical areas than those of Infomap-HSAs. Yet, Block Model-HSAs appear to be the poorest HSAs delineated not only because they are the fewest number of HSAs delineated, which corresponds to a wider and more discontinuous area, but also as their localization index is lower than $.5$ which raises concerns about the internal validity of the HSAs delineated. Lastly, Block Model-HSAs presented the highest variability regarding their localization index, conductance, and total number of discharges. Using the nested version of Block Model has not changed this results. Conversely, SLPA-HSAs are the highest number of HSAs but their localization index is typically less than $.7$, which is higher than that of Block Model-HSAs but still raises concerns regarding the internal validity of SLPA-HSAs.

The hospital discharge data and the Hospital-Patient Discharge Networks (HPDNs) inherently represent a flow between hospitals and patients and the apparent superior results achieved by Infomap may be related to the fact that, instead of modularity, the Infomap optimizes the map equation and thus takes into account the local flows emerging from the movements of random walkers trapped within the HPDN communities. Interestingly, Infomap-HSAs presented improved localization index and conductance over time.

\section{Conclusions}

Health Service Areas (HSAs) are meaningful units of analysis for improving the scientific basis of both clinical practice and policy decision making in the delivery of health care. The optimal delineation of HSAs is necessary to create not only more meaningful units of analysis, but also characterizing medical practices with greater accountability regarding their respective community needs and shared care practices. 

As HSAs shift from the Dartmouth approach towards a network-based approach, further work will be needed to establish a comprehensive methodology for network-based HSA delineation, which should include (i) a broader set of community detection algorithms, (ii) hospital discharge data from states other than California, and integration with other healthcare datasets of expenditures. Further work is still needed to establish a comprehensive methodology for this type of approach.

\bibliographystyle{plain} 
\bibliography{paper}

\end{document}